\documentclass{preprint}
\usepackage{graphicx}
\usepackage{booktabs}
\usepackage{multirow}
\usepackage{float}
\usepackage{array}

\title{On procedural urban digital twin generation and visualization of large scale data}
\author{S. Somanath, V. Naserentin, O. Eleftheriou, D. Sjölie, B. Stahre Wästberg\\ and Anders Logg}
\begin{document}
\maketitle


\begin{abstract}

 The desired outcome for urban digital twins is an automatically generated detailed 3D model of a building from aerial imagery, footprints, LiDAR, or a fusion of these. Such 3D models have applications in architecture, civil engineering, urban planning, construction, real estate, GIS, and many others. Further, the visualization of large-scale data in conjunction with the generated 3D models is often a recurring and resource-intensive task. However, a completely automated end-to-end workflow is complex, requiring many steps to achieve a high-quality visualization. Methods for building reconstruction approaches have come a long way from previously manual approaches to semi-automatic or automatic approaches. The next step after reconstructing buildings is visualizing the buildings and their context. Advances in real-time rendering using game engines have enabled the extension of building reconstruction methods to procedurally generated context generation. This paper aims to complement existing methods of 3D building generation. First, we present a literature review covering different options for procedurally generated context generation and visualization methods in-depth, focusing on workflows and data pipelines. Next, we present a semi-automated workflow that extends the building reconstruction pipeline to include procedural context generation (terrain and vegetation) using Unreal Engine and, finally, the integration of various types of large-scale urban analysis data for visualization. We conclude with a series of challenges faced in achieving such pipelines and the limitations of the current approach. The steps for a complete, end-to-end solution involve developing robust systems for building detection, rooftop recognition, and geometry generation and importing and visualizing data in the same 3D environment.
\end{abstract}

\section{Introduction} 

The urban environment is a complex and dynamic system that is constantly changing. The use of remote sensing data has become more accessible over the past decade, enabling regular monitoring and collection of urban data. Digital twins \cite{gil2020city}, virtual replicas of the physical world, have become a popular approach for simulating and analyzing the urban environment. However, traditional 3D modeling approaches for creating digital twins are often time-consuming and inflexible \cite{muller2006procedural}. In contrast, procedural model generation provides a more flexible workflow that automatically regenerates the built environment as data is updated.

In recent years, the concept of urban digital twins has emerged in the context of urban modeling and planning \cite{Ketzler}. Digital twins represent a city through its physical assets \cite{batty2018digital} and rich semantic data that represents city processes \cite{gil2020city}. They can potentially address the challenges of effective decision-making in complex systems like cities \cite{ham2020participatory}.

The VirtualCity@Chalmers outlines six characteristics a digital twin must possess - realistic, interactive, simulated, integrated, scalable, and open \cite{latino2019virtual}. Achieving these characteristics is critical to ensure the accuracy and reliability of urban digital twins. One of the most important aspects of achieving these characteristics is the process of visualization. Visualization can help to ensure that the digital twin is realistic, interactive, and scalable and can facilitate the effective dissemination of information within the digital twin.

This paper presents a workflow for the procedural generation of urban digital twins using multiple data sources, such as GIS data on building footprints, land use, and road networks in a game engine. The procedural approach provides a more flexible and efficient workflow that can be easily updated as the data is updated. This paper contributes to urban modeling and planning by providing a flexible and efficient workflow for generating urban digital twins. The proposed workflow and effective visualization techniques can improve decision-making in complex urban systems by providing a more accurate and reliable representation of the urban environment.

\section{Related work}

3D city modeling has seen a surge of activity from various research areas, each employing different methods and tools to create digital twins of cities from raw data. A common goal among researchers is to develop automated, robust, and efficient workflows to generate digital twins of cities with the highest level of detail possible. Ideally, the implementation should also be fast enough to allow for user interaction. In this chapter, we will explore different methods for efficient workflows in generating digital twins and the use of large-scale visualization to produce digital twins.

\subsection{Exploration of methods for efficient workflows in generating digital twins}

For the efficient production of digital twins, many researchers have focused on investigating methods and approaches concerning various aspects of development. Mu\~{n}umer Herrero et al. \cite{munumerherreroExploringExisting3d2022} compare different methodologies and software packages for exploring the possibility of creating various Levels of Detail (LoD) from a single data source. The methods used were TU Delft's 3Dfier, ArcGIS API for Python, and two different reconstruction processes by Poux~\cite{ph.d5StepGuideGenerate2021}, Polyfit ~\cite{nanPolyFitPolygonalSurface2017}, and RANSAC \cite{schnabelEfficientRANSACPointCloud2007}. However, none of these methods offered a complete solution, and the authors acknowledge that 3D generalization is essential. They also conclude that 3D models of different LoDs that are reconstructed in an automated way are of great interest.

\textbf{Visualisation models for urban planning and LoD's}: Regarding landscape modeling, Wang \cite{Wang2022} discusses using 3D landscapes in landscape design while comparing landscape modeling methods. According to the author, people tend to automatically make decisions based on physiological responses and engage more with nature when visualizing an interactive landscape. The paper concludes that introducing 3D technology is critical in optimizing urban planning and management. 

At the same time, Coors et al. ~\cite{coorsConceptQualityManagement2020} attempt to specify the application-specific requirements for 3D urban models and distinguish between simulation-able and those just for visualization purposes. The study results show no one-size-fits-all approach to LoD and geometry attributes, the same as García-Sánchez et al. ~\cite{garcia-sanchezIMPACTLEVELDETAIL2021}. They indicate that the oversimplification of 3D city model geometries can affect the results of Computational Fluid Dynamics (CFD) for wind flows. The authors conclude that models at different LoDs (1.3 and 2.2) and semantic configurations (areas of water and vegetation) show significant differences in wind patterns in built environments. They advocate for higher LoD (2.2) and semantics for better results. 

Deininger et al. ~\cite{deiningerContinuousSemiAutomatedWorkflow2020} addressed the challenges of generating CFD-ready models, highlighting the need for accurate and efficient models that can simulate fluid flows through urban environments. They discuss the various techniques for generating such models, including meshing techniques, data assimilation methods, and machine learning algorithms.
Kolbe et al. ~\cite{kolbeSemantic3DCity2021} also explored the differences between semantic 3D models and Building Information Models (BIM) on the urban scale. They conclude that the achievable and manageable data quality of urban models is limited by the data collection processes and the employed standards concerning the data modeling frameworks and data exchange capabilities. These findings highlight the need for standardization in 3D city modeling to enable interoperability and data sharing between different systems and applications.

\textbf{Data sources for 3D city models}: Common data sets used to create city models are building footprints and digital elevation models. Singla et al.~\cite{singlaNovelApproachGeneration2021} presented a cost-effective approach to generating 3D city models from digital elevation models and open street maps. In contrast, Girindran et al. ~\cite{girindranReliableGeneration3D2020} proposed another cost-effective and scalable methodology for generating 3D city models that relies on open-source 2D building data from OpenStreetMap and open satellite-based elevation datasets. The approach allows for increased accuracy should higher-resolution data become available. This methodology targets areas where free 3D building data is unavailable and is largely automated, making it an attractive option for cities with limited budgets.

Other data sources used to create city models include stereo aerial images. Pepe et al. .~\cite{pepeNovelMethodBased2021} presented a 3D city model generation method using high-resolution stereo imagery as input. The proposed approach generates high-quality 3D models while overcoming some of the limitations of traditional photogrammetry-based methods. 

With high-resolution satellite images and LiDAR scanning becoming more available, these new data sources are more commonly used for creating city models. Buyukdemircioglu et al. ~\cite{buyukdemirciogluReconstructionEfficientVisualization2020} attempted to combine orthophotos and high-resolution terrain models to serve as the base data for a 3D GIS environment. They showcased the different stages involved in building LoD 2 and 3 models and their potential extension to LoD 4 in the (near) future. 

Dollner et al. ~\cite{dollnerContinuousLevelofdetailModeling2005} leveraged the LandXplorer system as an implementation platform to create continuous LoDs. Their approach enabled buildings to offer a continuous LoD, which is significant for various urban planning stages. Ortega et al. ~\cite{ortegaGenerating3DCity2021} present a method for creating LoD 2 3D city models from LiDAR and cadastral building footprint data. They also classified the roofs of the buildings into one of five possible categories using purely geometrical criteria to enable the creation of higher LoD models.

Exploring other city-scale simulations, Katal et al. ~\cite{katalUrbanBuildingEnergy2022} proposed a workflow that integrates urban microclimate and building energy models to improve accuracy in 3D city models. Non-geometric building parameters and microclimate data were included after creating the 3D model of buildings to capture the two-way interaction between buildings and microclimate. By coupling CityFFD and CityBEM, significant changes in calculated building energy consumption are achieved.

\textbf{Towards procedural workflows}: Recently, Chen et al.~\cite{chenReconstructingCompactBuilding2022} demonstrated a promising method that uses deep neural networks and Markov random fields to reconstruct high-fidelity 3D city models. They compared their work with other methods and found that their new approach offers better overall results and is fast enough for user interaction. Vitalis et al. ~\cite{vitalisAPPLYINGVERSIONINGMULTILOD2022} propose a framework for managing the evolution of multi-LoD datasets to keep track of the temporal aspect of different LoDs. 

Meanwhile, Halim et al. ~\cite{halimDeveloping3DCity2021} highlight the background and purpose of SmartKADASTER's 3D city model database. They outline the approaches used to develop a platform for cadastral survey information with 3D visualization, migration to CityGML, and PostgreSQL database providing representations at different LoDs. For more comprehensive reviews of state-of-the-art 3D city model generation of varying LoD, readers can refer to ~\cite{Naserentin2022,naserentinCombiningOpenSource2022a,chenReconstructingCompactBuilding2022,  munumerherreroExploringExisting3d2022}.

In summary, researchers have been making strides toward developing automated, robust, and efficient workflows for creating digital twins of cities from raw data. However, there is no one-size-fits-all approach to LoD and geometry attributes. García-Sánchez et al.'s work highlights that the oversimplification of 3D city model geometries affects CFD results for wind flows, emphasizing the need for higher LoD and semantics. Singla et al. and Coors et al.'s work provide cost-effective approaches to generate 3D city models and specify application-specific requirements for 3D urban models, respectively. The study results also indicate that automated 3D models capable of different LoDs are of great interest for both visualization and simulation purposes.

\subsection{Large scale data visualization in the production of digital twins}
Apart from 3D city modeling, large-scale data visualization is a crucial component in developing digital twins for cities. With the growing use of Internet of Things (IoT) devices and smart technologies, there is an enormous amount of data generated by various sources in a city, ranging from traffic patterns, energy consumption, air quality, and more. The challenge is to make sense of this data, and one way to achieve this is through effective data visualization techniques\cite{ketzlerDigitalTwinsCities2020b}.

Data visualization is the process of representing data graphically, enabling people to quickly and easily understand complex information. In the context of digital twins for cities, large-scale data visualization can provide valuable insights into the functioning of a city, such as identifying patterns, trends, and anomalies\cite{UrbanTrafficSpeedEstimation}. It can help city planners and policymakers to make informed decisions about various aspects of the city's infrastructure, including transportation, energy, and environmental management \cite{EULUC}.

Moreover, large-scale data visualization can facilitate effective communication among different stakeholders, such as citizens, businesses, and government agencies. By presenting data in an accessible and engaging manner, it can help build awareness and support for various initiatives to improve the quality of life in cities. Overall, research on large-scale data visualization is critical for the development of digital twins for cities, as it provides a powerful tool for understanding and managing the complex urban systems of today and tomorrow \cite{gil2020city}. 

\textbf{Types of data visualization}: In a recent study, a novel interactive system for fast queries over time series is presented \cite{VisComputerGraphics}. The developed system can perform efficient line queries and density field computations while providing fast rendering and interactive exploration through the displayed data. \cite{Time-Evolving-3DCityModels} propose a method to model, deliver and visualize the evolution of cities on the web. The authors developed a generic conceptual model with a formalization of the temporal dimension of cities. Afterward, researchers proposed a logical model for time-evolving 3D city models. The authors suggest that their visualization platform drastically improves temporal navigation. 

On a lower computer architectural level, a GPU-based pipeline ray casting method was proposed to visualize urban-scale pipelines as a part of a virtual globe \cite{wu2019gpu}. The results of the approach indicated that the proposed visualization method meets the criteria for multiscale visualization of urban pipelines in a virtual globe which is of great importance to urban infrastructure development.

\textbf{GPS and mobility data}: By leveraging trajectory data generated by vehicles with GPS-capable smartphones, researchers performed a network-wide traffic speed estimation \cite{UrbanTrafficSpeedEstimation}. The proposed system is capable of generating congestion maps that can visualize traffic dynamics. By visualizing the data, the system became vital as it enabled experts to continuously monitor and estimate urban traffic conditions, which ultimately improved the overall traffic management process.

In another study, by visualizing a novel analytical method of bike-sharing mobility \cite{Kon2021}, the authors obtained relevant mobility flows across specific urban areas. The data visualization can aid public authorities and city planners in making data-driven planning decisions. The authors conducted an assessment of their system with field experts. They concluded that the proposed system is easy to use and can be leveraged to make decisions regarding the cycling infrastructure of cities that provide bike-sharing.

To help alleviate urban traffic congestion, a deep learning multi-block hybrid model for bike-sharing using visual analysis about spatial-temporal characteristics of GPS data in Shanghai was proposed \cite{Bike-Sharing-Supply-Demand-Prediction}. The authors rendered the supply-demand forecasting of the bike-sharing system. By capturing spatiotemporal characteristics of multi-source data, they could effectively predict and optimize supply-demand gaps which are vital in rebalancing the bike-sharing system in the whole city. 

Jiang et al. leveraged large-scale vehicle mobility data to understand urban structures and crowd dynamics better \cite{Zhihan}. The authors combined visualization systems with a data-driven framework that senses urban structures and dynamics from large-scale vehicle mobility data. Additionally, they included an anomaly detection algorithm to correlate irregular traffic patterns with urban social and emergency events. The authors conclude that their framework effectively senses urban structures and crowd dynamics, which are crucial for urban planning and city management.

\textbf{Land use and energy data}: Apart from different traffic scenarios, data visualization is also used for visualizing other areas in a digital twin. When mapping essential land use categories (EULUC), a review by \cite{EULUC} presents the progress, challenges, and opportunities using geospatial big data. The authors indicate that land use information is essential for landscape design, health promotion, environmental management, urban planning, and biodiversity conservation. Additionally, the authors propose various future opportunities to achieve multiscale EULUC mapping research.

A novel visual analysis system ElectricVIS for urban power supply situations in a city was proposed by\cite{Lu2020}. The system can be leveraged to interactively analyze and visualize large-scale urban power supply data. Using time patterns and different visual views, ElectricVIS aids power experts in detecting the cause of anomalous data. After performing user evaluation, the authors concluded that experts spent less time overall when using ElectricVIS compared to a traditional view system to make estimations and make informed decisions.

\textbf{Interactivity in data visualization}: Using the right visualization workflow can have real-world benefits by providing valuable insights to decision-makers at the city level in complex scenarios \cite{Deng2021-ms}. Deng et al. proposed a visual analytics system combining inference models with interactive visualizations in order to allow analysts to detect and interpret cascading patterns in spatiotemporal context \cite{Deng2021-ms}. After conducting two case studies with field experts, the authors concluded their system could be leveraged to reduce the time-consuming process of identifying spatial cascades. Additionally, based on expert feedback, the system is deemed applicable to large-scale urban data.

An empirical investigation is presented by Gardony et al. in \cite{EffectiveAR-Geo-Vis} regarding how user interaction in AR systems can affect users. The users were given an interactive 3D urban environment to learn an embedded route between two locations. The study indicated that users who used the city model to gain an overhead followed the designated route. On the other hand, users who consistently interacted with the model could unexpectedly and efficiently return to the router's origin. The paper concludes that depending on the task at hand, more research should be done on whether to provide task-relevant views or fully dynamic interaction to users.

\textbf{Visualisation platforms}: In \cite{Intelligent-Transportation-Systems}, a web-based traffic emulator is presented. The application uses collected data from roadside sensors to visualize near-real-time and historical traffic flows. User evaluations indicated that visualizations of traffic flow with LoD techniques could reveal traffic dynamics of emulated traffic from the microscopic to the macroscopic scale.

More recently, real-time rendering using game engines has gained popularity in visualizing 3D city models. Lee et al. propose a planetary-scale geospatial open platform using the Unity3D game engine in \cite{Planetary-Scale-Geospatial-Open-Platform}. To generate objects in the 3D world, they used VWorld's geospatial data. Their platform can visualize large-capacity geospatial data in real-time, while the authors believe the proposed platform meets the needs of various 3D geospatial applications.

In addition to technology developed for games and film, techniques from these domains have also been adopted to generate effective visualization. A camera-shot design approach to tracking evacuation changes and correlations in earthquake evacuation is proposed by Q. Li et al. in \cite{SEEVis}. After several different case studies, the authors conclude that their system is efficient in helping experts and users alike to gain a better insight into earthquake evacuation by assisting them in developing a comprehensive understanding of the situation.

\section{Materials and Methods}

The methodology described in this paper consists of two parts, a - \textit{World Creation} and b - \textit{Data Visualisation}. World Creation uses available GIS data to generate a 3D virtual city model (VCM) and visualize the data using Unreal Engine. Data visualization is concerned with visualizing raw data spanning a large spatial extent. This step aims to achieve large-scale visualization while maintaining the procedural nature of the workflow (see Figure \ref{flowchartCombined} ). The following sections describe the methods and datasets used in each step.
    \begin{figure}[!htb]
        \centering
        \includegraphics[width = 0.9\textwidth]{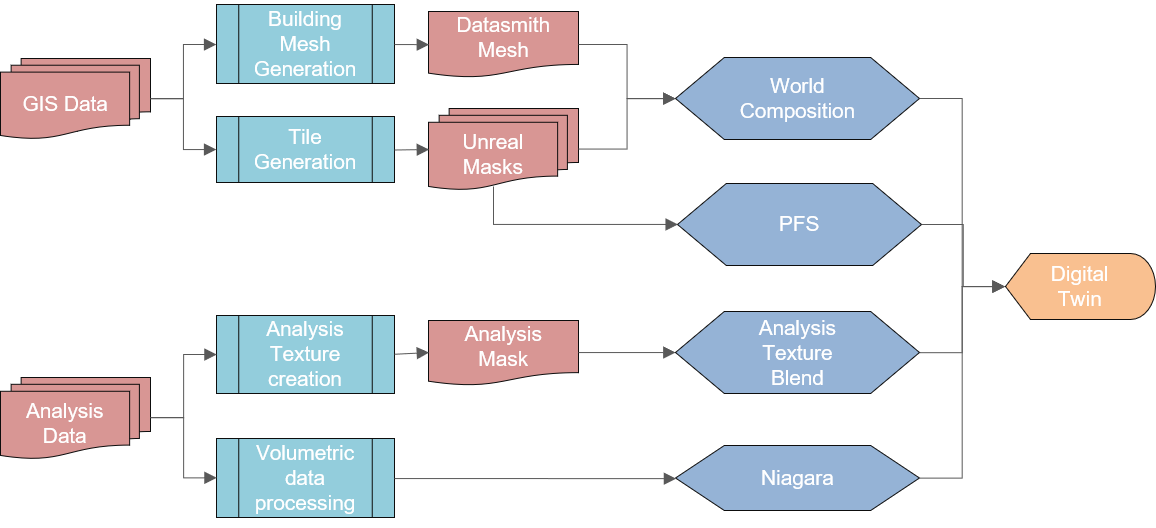}
        \caption{Combined workflow for World creation and Data visualization.}
    \end{figure}
    \label{flowchartCombined}

\subsection{World Creation}

\begin{table}[!htb]
\caption{Used datasets from Lantmäteriet.}
\resizebox{\textwidth}{!}{%
\begin{tabular}{@{}llll@{}}
\toprule
\textbf{Swedish   name} & \textbf{English name} & \textbf{Description}                       & \textbf{Format} \\ \midrule
Bebyggelse              & Property map          & Building footprints as closed polygons     & SHP             \\
Kommunikation           & Transport networks    & Transportation network as center lines     & SHP             \\
Markdata                & Land data             & Land use classification as closed polygons & SHP             \\
Höjddata                & Elevation data        & Ground elevation data as raster images      & TIF             \\
Laserdata               & Laser data            & LiDAR scan data as point clouds            & LAZ             \\ \bottomrule
\end{tabular}%
}

\label{tab:dataset}
\end{table}

\begin{figure}[!htb]
        \centering
        \includegraphics[width = 1\textwidth]{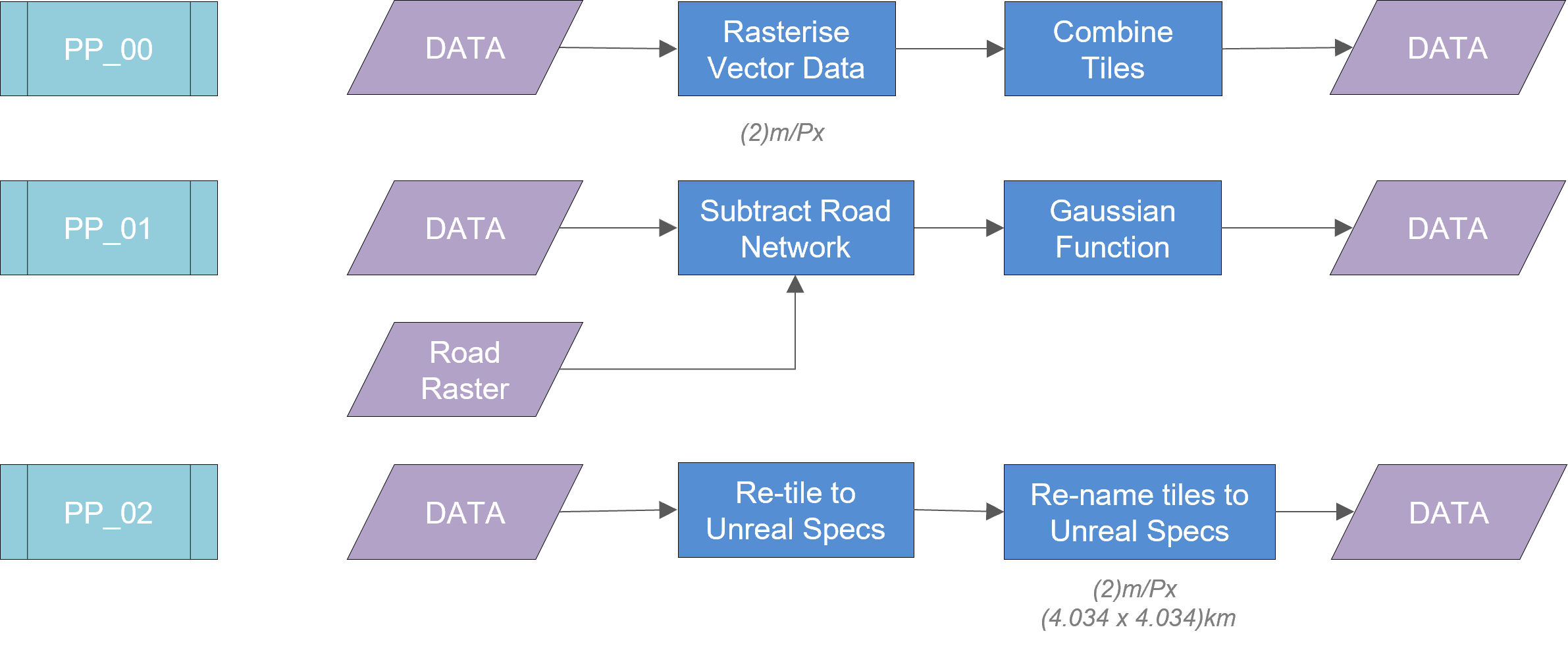}
        \caption{Pre-defined processes used in world generation workflow.}
        \label{PredefinedProcess}
    \end{figure}

 For world-creation, we use GIS data provided by Lantmäteriet \cite{LM2022} (see Table \ref{tab:dataset}), the Department of Land Survey, to procedurally build the natural and built environment of the selected region. For the natural environment, the features we choose for this paper are geographical features and vegetation, whereas the built environment consists of existing buildings and roads. However, this workflow can be extended to finer details of the built environment, such as road markings, traffic signage, and street furniture.
We use three pre-defined processes repeatedly throughout the world generation workflow. Figure \ref{PredefinedProcess} illustrates the workflow of these pre-defined processes. The pre-defined processes are used to rasterize vector GIS data and combine them into a single raster image, subtract the buffered road networks from the raster layers, and generate Unreal Engine-compatible raster tiles from the processed raster image.

\subsubsection{Terrain}

\begin{figure}[!htb]
        \centering
        \includegraphics[width = 1\textwidth]{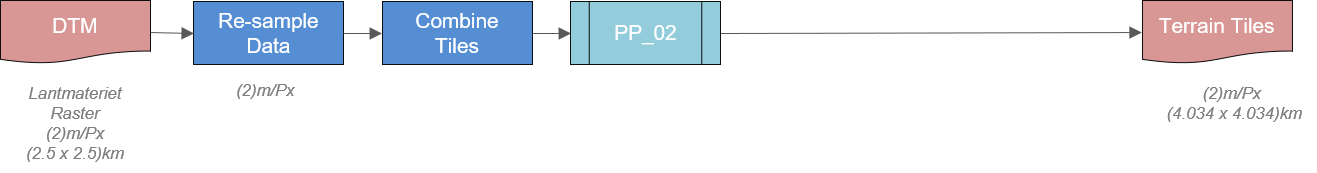}
        \caption{Workflow to generate terrain tiles.}
        \label{FlowchartTerrain}
    \end{figure}

Lantmäteriet provides a Digital Terrain Model (DTM) at a two-meter-per-pixel resolution of 6.25 square kilometer tiles. To ensure the workflow is procedural, we introduce a post-processing step that re-samples the incoming DTM raster to a pre-determined cell spacing. The DTM is then combined into a single raster and further re-tiled into specific dimensions to ensure the maximum area while minimizing the number of tiles within the game engine \footnote{\url{https://docs.unrealengine.com/5.0/en-US/landscape-technical-guide-in-unreal-engine/}}. The final step in creating the terrain tiles is to ensure the tiles are systematically named for the game engine to identify the tiling layout.
Once the terrain tiles are generated, they can be imported into the game engine using the World Composition feature. Figure \ref{FlowchartTerrain} shows a flowchart of the different steps used to generate the terrain masks.

\subsubsection{Roads}

\begin{figure}[!htb]
        \centering
        \includegraphics[width = 1\textwidth]{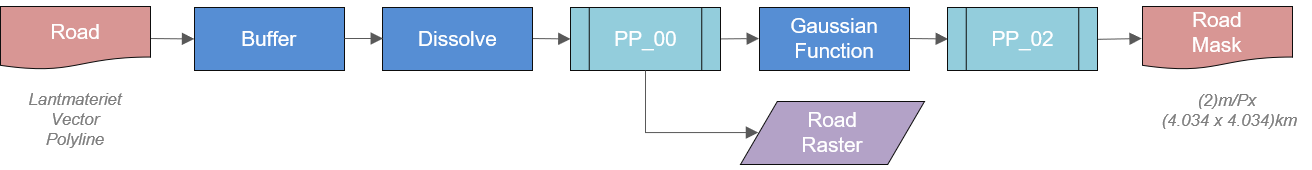}
        \caption{Workflow to generate road masks.}
        \label{FlowchartRoads}
    \end{figure}

Lantmäteriet and Trafikverket provide road data as road center-line polygonal vector data. The vector data consists of twenty road classes, such as tunnels, thoroughfares, public roads, roads under construction, and a range of road widths. We simplify the roads into four classes and assign fixed road widths to each class. The center-line data is then post-processed to form landscape masks for the game engine. We first perform a spatial buffer across all centreline polygons per road class and then dissolve the resulting polygon offset to form the road boundary regions. We then rasterize, tile, and rename the road tiles as shown in Figure \ref{FlowchartRoads}.

\subsubsection{Vegetation}

\begin{figure}[!htb]
        \centering
        \includegraphics[width = 1\textwidth]{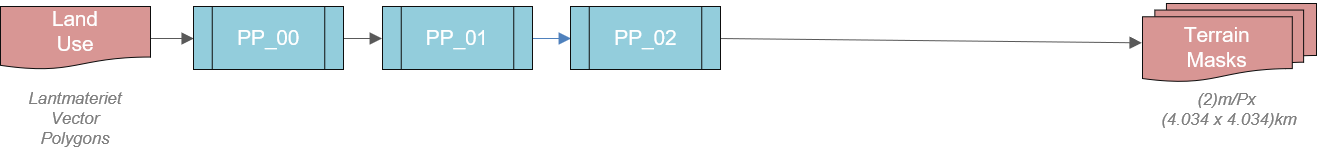}
        \caption{Workflow to generate terrain masks.}
        \label{FlowchartVegetation}
    \end{figure}

Lantmäteriet  provides data regarding land-use classification in the form of vector polygons. The data consists of 13 natural and human-made land-use classes: Water, coniferous forests, low built-up, high built-up, and industrial areas. For simplicity, we reduce the land-use classes to five - Water, forest, farm, and urban.
For each land-use class, the vector data is first rasterized with binary cell values (0 representing the absence of the land-use class and 1 representing its presence) to the predefined cell spacing as the terrain tiles. The process of tiling and renaming output tiles is carried out as shown in Figure \ref{FlowchartVegetation}.

\subsubsection{Buildings}

\begin{figure}
        \centering
        \includegraphics[width = 1\textwidth]{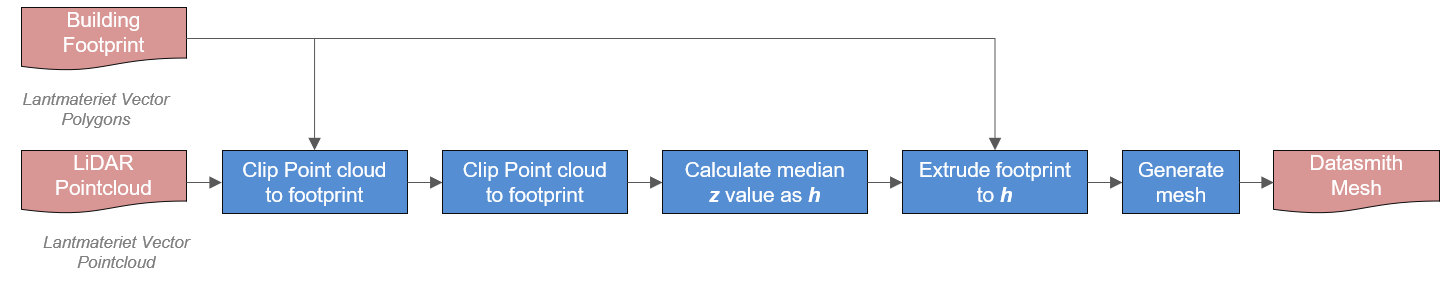}
        \caption{Workflow to generate building meshes.}
        \label{FlowchartBuildings}
    \end{figure}

The building meshes are generated using two data layers provided by Lantmäteriet, the building footprints and a LiDAR point cloud. First, we determine the mean height of the buildings by averaging the values of points in the z dimension above a building's footprint. A LoD1  building mesh is generated by extruding the building footprints to their respective heights. The buildings are then repositioned in the z dimension to meet the terrain and generate a DataSmith file from the building meshes and import them into the game engine using the software Feature Manipulation Engine (FME) as shown in Figure \ref{FlowchartBuildings}

\subsection{Unreal Engine workflow}
\subsubsection{Integration of landscape tiles}

\begin{figure}[!htb]
        \centering
        \includegraphics[width = 0.6\textwidth]{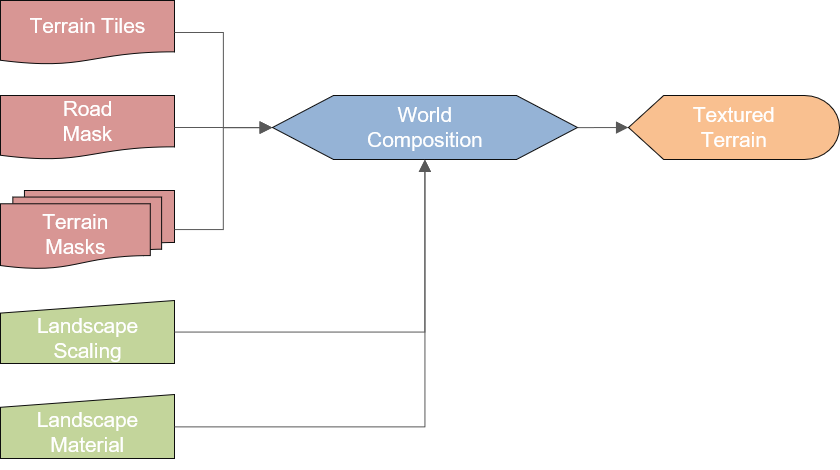}
        \caption{Combined workflow to generate textured terrain.}
        \label{figure_combined_workflow}
    \end{figure}

The final step in preparing the landscape tiles is to ensure that the different layers are incorporated into a seamless composition. There are two issues that we address in this step. First - the land-use boundaries are generated irrespective of the road networks; this causes an overlap of binary raster values at the intersection of a land-use class and a road segment; second - the process of rasterizing vector data results in a steep drop in the cell values, which is presented as pixelated boundaries at the edge of the boundaries.
Using FME, first, we subtract the raster data for the land-use layers with the road network to avoid any overlap; then, we apply a raster convolution filter across the layers using the Gaussian function, ensuring a gradual falloff in values at the boundaries of the layers.
To complete the integration process, we import the terrain mask into the game engine using the World Composition feature, provide the scaling factors in the x,y, and z dimensions and assign the subsequent landscape masks to a procedural landscape material as shown in Figure \ref{figure_combined_workflow}. 

\subsubsection{Procedural Landscape Material}
In its material editor, Unreal Engine offers a "Landscape Layer Blend" node. The node enables us to blend together multiple textures and materials linked to the landscape masks generated in the World creation step. A layer is added for each class of landscape material (farm, forest, water, etc.), and a texture is assigned to it.

\subsection{Data visualisation}
\begin{figure}[!htb]
        \centering
        \includegraphics[width = 1\textwidth]{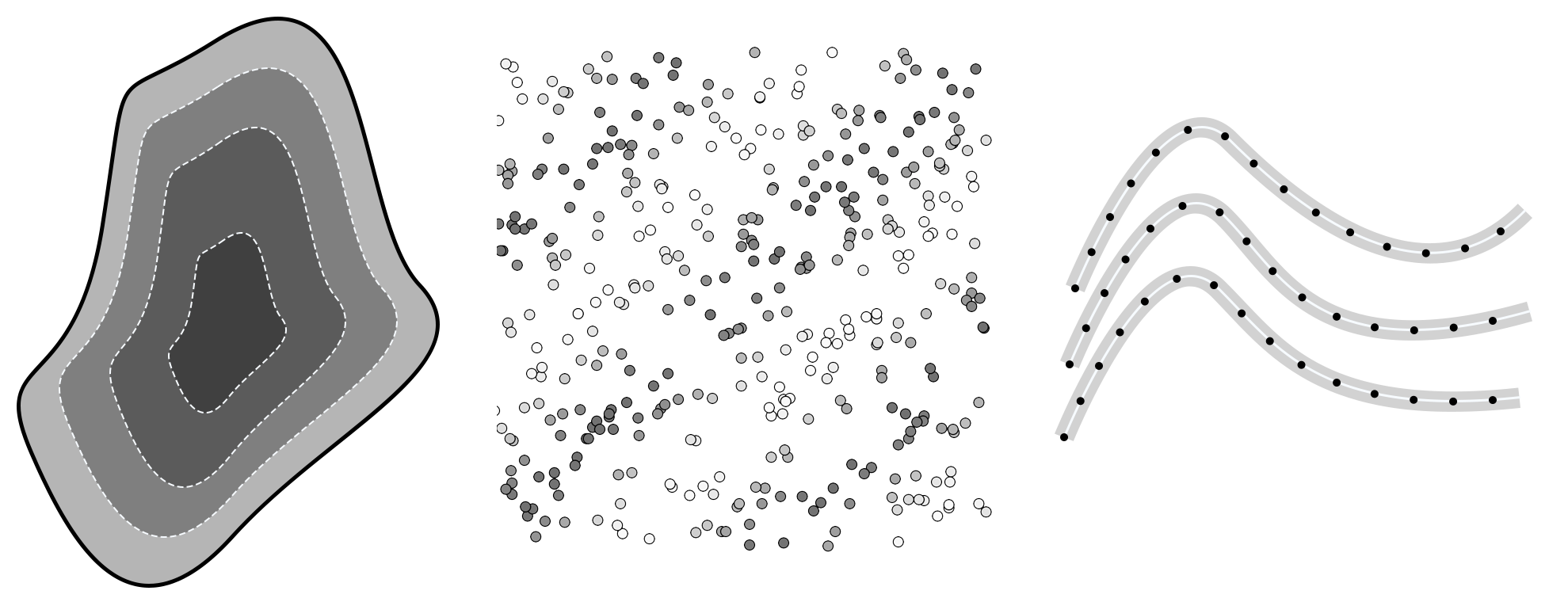}
        \caption{Illustration of the three data visualizations presented in this paper: isoline, volumetric, and streamlines.}
        \label{figure_illustration}
    \end{figure}

Visualization, virtual experimentation, and test-bedding are some of the key applications of VCMs\cite{Ketzler}. The data to be visualized is produced by sensors, analysis, and simulations in various data types. The following section outlines visualization methods for isoline, volumetric, and 3D streamline data (see Figure \ref{figure_illustration}). The isoline data is visualized by overlaying a scaled color value on the terrain texture. Volumetric data is visualized as volumetric particles of scaled color values, and 3D Streamlines are visualized as either static or moving streamlines of particles also of a scaled color value. The data used for the different visualization were generated from wind and noise simulations conducted by other project partners.

\subsubsection{Isoline Data}
 Results from analyses conducted over a large area, such as noise and air quality, are available as 2D isolines from other project partners. The closed curves in the data set represent a region with a constant value, and adjacent curves represent a fixed change in the value. The isoline data is first converted to a raster image using FME, where we pack our data into a single grayscale image to have normalized values. We then extended the procedural landscape material to visualize data on top of the terrain, allowing us to paint pixels in certain areas that matched our data.  Considering that our data was packed into a single texture and each terrain tile had the same material instance applied, we introduced a custom material function in the editor that allows us to use different data textures with various dimensions and colormaps. 

First, we scaled the data texture to match the available terrain data. Then, in the material code, we accessed each pixel of the data and read its normalized value to map it to its respective color from the assigned color map. Afterwards, we performed a linear interpolation between the base terrain layer (containing the existing materials of the environment, such as water, grass, and dirt) and the visualized data texture, thus allowing us to display colored data on top of the terrain while showcasing the natural environment in the areas where no data was available in the data texture.

\subsubsection{Volumetric data}
\begin{figure}[!htb]
        \centering
        \includegraphics[width = 1\textwidth]{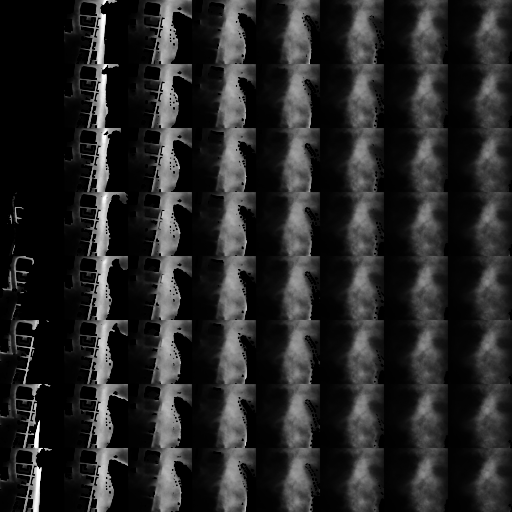}
        \caption{Data texture with 64 (8x8) z-levels. The contrast has been enhanced here for illustrative purposes.}
        \label{figDataTexture}
    \end{figure}

To visualize volumetric data available throughout a 3D space, such as air pollution or noise levels, we again employed the strategy of encoding the data into textures (\textit{data textures}) (see Figure \ref{figDataTexture}) that can be used for real-time visualization in Unreal Engine. These textures are then used to, e.g., create dynamic cut-planes using Materials coded in Unreal Engine to look up the data dynamically based on the current 3D position on the surface. As these textures fold 3D data into a 2D texture, particular care must be taken to interpolate data along the z-axis (up/down in Unreal Engine). Data from the same xy-coordinate is sampled from two places in the data texture, corresponding to the two closest z-coordinates in the data. Then this data is interpolated to reflect the exact z-coordinate that should be sampled. The same data textures are also used as the basis for particle visualizations using Niagara, changing color, size, lifetime, etc., of particles depending on the data corresponding to the particle's position in 3D.

The support for Python scripting in the Unreal Engine Editor provides an efficient workflow for generating the data textures. Python makes it easy to read and prepare most data types using commonly available Python libraries, such as numpy and automatically imports these into native Unreal Engine Textures.

\paragraph{Particle streamlines}
We have explored two different strategies for generating 3D streamlines, showing, e.g., air flows in the 3D space around and between buildings.

First, we used C++ to generate procedural meshes, creating a colored tube for each streamline in 3D. This visualization gave high fidelity and is suitable for a limited number of static streamlines but did not allow for performant visualization of a large number of streamlines.

To enable the visualization of several thousands of streamlines, we used the Niagara particle system integrated into Unreal Engine. By encoding streamlined data into data textures, as described above, the locations of streamlined segments were made accessible to the GPU using the Niagara particle system in Unreal Engine. This data generated one particle per segment of the streamlines and placed them with the correct positions and orientations to make up streamlines. With this approach, we can get real-time performance with 10,000+ streamlines with up to 1000 line segments per streamline, corresponding to several millions of line segments.

One challenge with this approach was that the position data saved into the texture needed higher precision than what is commonly used in textures for computer graphics. To properly encode the data, 32-bit textures were required. While this is supported in many key places along the pipeline (e.g., in Python and Niagara), it was not supported in the standard functions for importing textures into Unreal Engine at the time of this implementation. As such, we created and imported two 16-bit textures and merged them into one 32-bit texture using \textit{RenderTargets} that currently support 32-bit texture data. This 32-bit RenderTarget texture can be accessed directly from Niagara, providing the required data precision.

\paragraph{Procedural meshed streamlines}
To create procedural streamlines, we parse the data for each streamline, a collection of points. Once we have parsed all the points for a given streamline, we generate a cylinder-like mesh around each point. The mesh generation process requires the following steps:
\begin{enumerate}
    \item Generating vertices around each parsed point of the streamline
    \item Creating triangles between the generated vertices
    \item Connecting sequential vertices that belong to different points
\end{enumerate}
To demonstrate how the algorithm works, consider the following case. Assume that we have parsed point O from our data. As a first step based on the description above, we need to generate vertices around that point. To do that, we have exposed a value that can be modified in Unreal Engine, called \textit{CapVertices}, which is the number of vertices that will be generated around point O. Using the polar coordinate system, we can create an initial vector $\Vec{OA}$ which we are going to rotate \textit{360/CapVertices} degrees for \textit{CapVertices} times around X axis to create the required vertices for the parsed point:
    \begin{figure}[!htb]
        \centering
        \includegraphics[width = 0.45\textwidth]{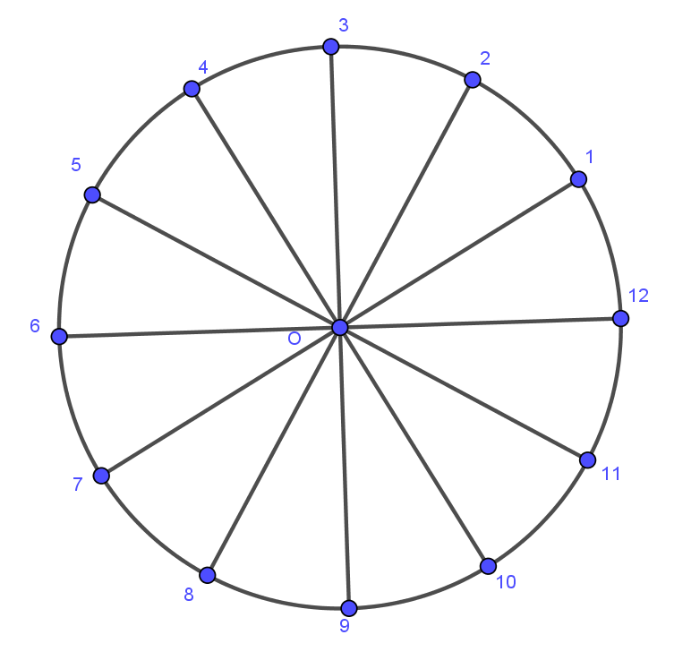}
        \caption{Generated vertices around point O.}
        \label{fig_streamlines}
    \end{figure}
\pagebreak

\begin{figure}
        \centering
        \includegraphics[width = 0.85\textwidth]{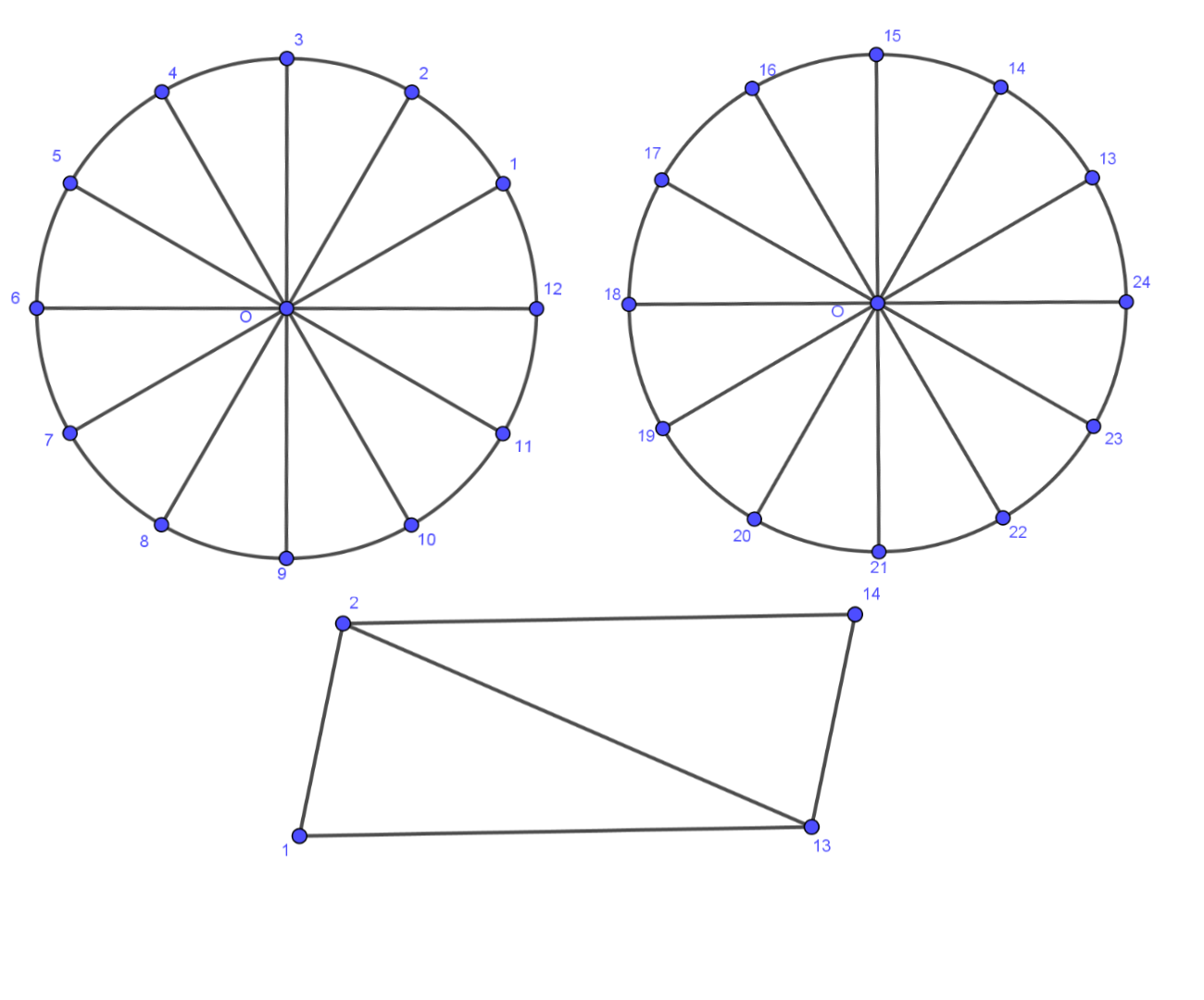}
        \caption{Upper images: Generated vertices around the first and second points of streamline. Lower image: A flattened display of connected vertices of two sequential points.} \label{streamline_shapes}
    \end{figure}

Once the vertices have been generated, the next step of the process is to create the required triangles to connect them. Considering Figure \ref{fig_streamlines}, we generate a cylinder cap for the parsed point by connecting every three vertices in a counterclockwise fashion. This process is repeated for all the parsed points of a single streamline. For the mesh to look streamlined, we must connect the generated vertices belonging to different points. Figure \ref{streamline_shapes} displays the generated vertices for the \textit{first} and \textit{second} points of a streamline, respectively. Then, to create a cylinder-looking mesh between the two streamline caps, we generate a parallelogram for vertices \textit{1,2,13 and 14} (displayed in the lower section on Figure \ref{streamline_shapes}.

For the mesh to appear smooth and consistent, we must connect vertices \textit{2,13 and 1} and \textit{2,14 and 13} in a counterclockwise fashion. By repeating this process for all the generated vertices around all points in each streamline, we create a single cylinder-like mesh for the whole streamline.


\section{Results}
This section presents the results of the World creation and data visualization workflows. The results from the world creation workflow consist of asset generation in FME and the time required to generate and load the assets. Then, we present the results of the Unreal Engine workflow. In the next section, Data visualization, we present the results of each visualization type - isoline data, volumetric data, and streamline data.
\subsection{World Creation}
\subsubsection{Generating Assets}
The workflow for generating assets is implemented in the FME software. This is an automated pipeline where the only input required is the location of the raw GIS data.
Table \ref{tab:timing} shows the time required for each step to be completed in FME and, similarly, the time required to load the assets into Unreal Engine. Steps requiring manual inputs, such as configuring the import locations and settings into Unreal Engine and entering the values to align the building meshes in Unreal Engine, are left blank.
\begin{table}[H]
\centering
\caption{Time required to procedurally generate the 3D environment.}
\begin{tabular}{@{}cclll@{}}
\toprule
\multicolumn{1}{l}{}             & \multicolumn{1}{l}{}           & Process                          & Time (s) & Software \\ \midrule
\multirow{9}{*}{Generate Assets} & \multirow{9}{*}{23.43 Minutes} & Writing Buildings & 513.1    & FME      \\
 &  & Writing Height maps       & 42.6  & FME    \\
 &  & Writing Mask - Water     & 8     & FME    \\
 &  & Writing Mask - Forest    & 14.1  & FME    \\
 &  & Writing Mask - Farm      & 264.2 & FME    \\
 &  & Writing Mask - Urban     & 113.8 & FME    \\
 &  & Writing Mask - Open      & 151.1 & FME    \\
 &  & Writing Mask - Road      & 165.8 & FME    \\
 &  & Writing Analysis Overlay & 132.8 & FME    \\
\cmidrule(l){3-5} 
\multirow{5}{*}{Load Assets}     & \multirow{5}{*}{6.11 Minutes}  & Unreal import config             & -        & UE   \\
 &  & World Composition        & 205.7 & UE \\
 &  & Import Building Mesh     & 152.5 & UE \\
 &  & Align Meshes             & -     & UE \\
 &  & Foliage                  & 8.42  & UE \\ \cmidrule(l){3-5} 
\end{tabular}
\label{tab:timing}
\end{table}
\subsubsection{Unreal Engine workflow}
    \begin{figure}[!htb]
        \centering
        \includegraphics[width = 1\textwidth]{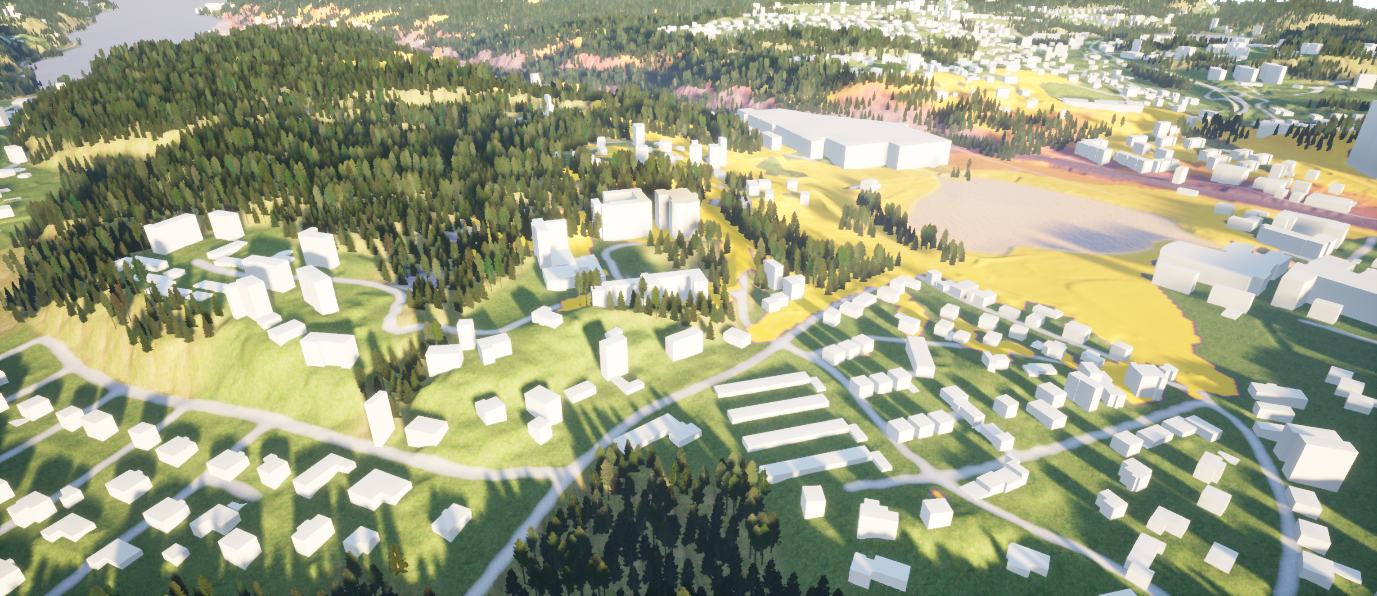}
        \caption{Aerial view of the final 3D model - Unreal Engine.}
    \end{figure}

    \begin{figure}
        \centering
        \includegraphics[width = 1\textwidth]{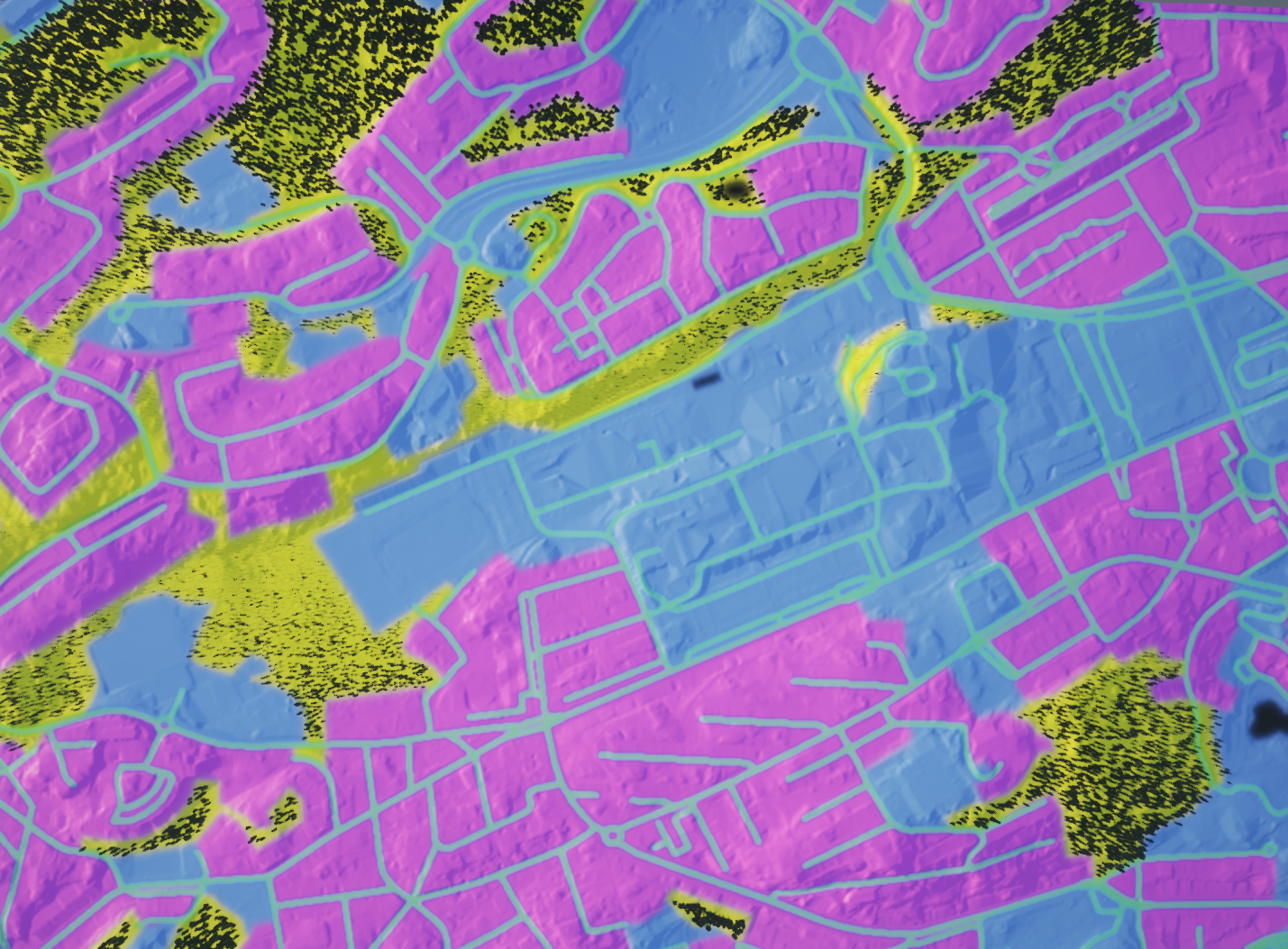}
        \caption{Aerial view of terrain model with landscape masks applied - Unreal Engine.}
    \end{figure}

    \begin{figure}
        \centering
        \includegraphics[width = 1\textwidth]{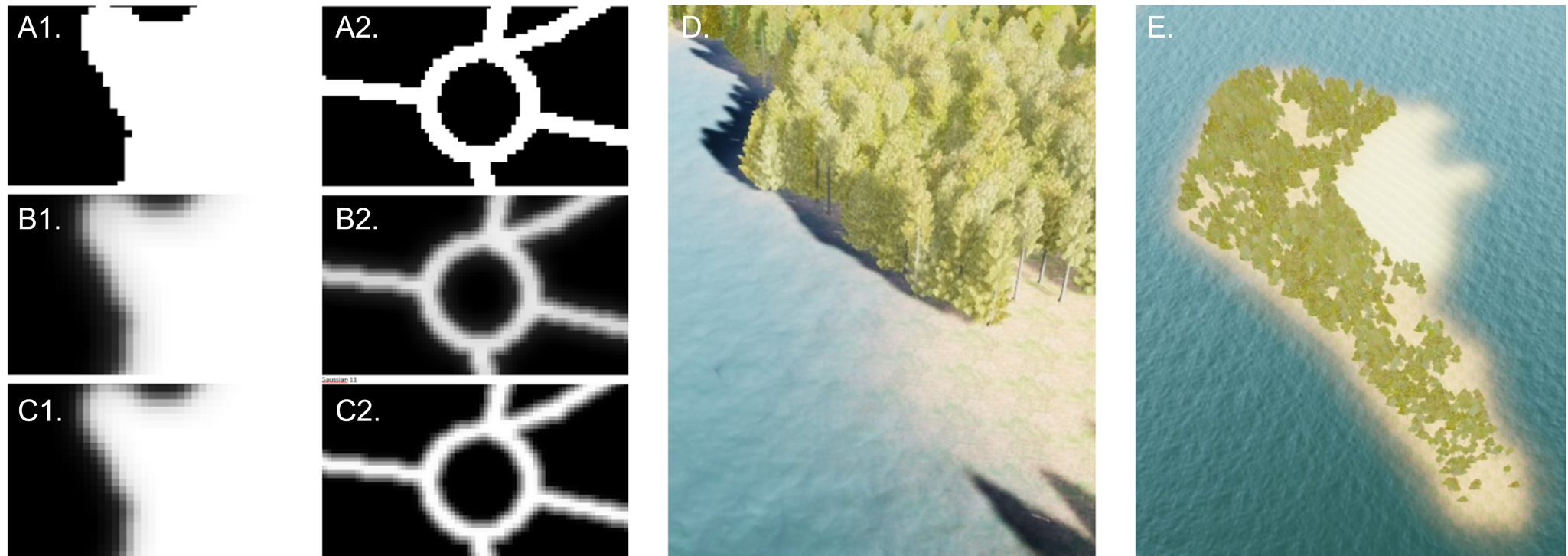}
        \caption{(Left) Results of convolution filters on the landscape masks. (Middle), (Right) Result of convolution on landscape masks - Unreal Engine.}
        \label{result_convolution}
    \end{figure}
Lantmäteriet provides the data containing the boundary for each land-use category. These categories include land use types such as roads, water, forest, farm, urban, and open land. Since the Lantmäteriet data is available as vector data, it must be first converted into raster data before Unreal Engine can use it. Unreal Engine uses these black and white masks as raster images to identify where the different materials must be applied within the 3D model (black representing the absence of a category and white representing the presence of it). One of the problems we encounter in this process is the interaction between the different land use types at the edges of the boundary. The edges are sharp and do not blend into one another naturally. We use a Gaussian function to smooth the edges naturally to solve this. The softening of the edges is achieved using convolution, where a matrix operation is performed on an array of pixels according to a set of pre-defined weights. Figure \ref{result_convolution} – a1 and a2 show a land use mask for a region containing forests and roads, respectively, with no filtering. Figure \ref{result_convolution} – b and c show the results of different weights provided to the Gaussian function.
Figure \ref{result_convolution} - d and e show the results of the layer blending once the processed masks are loaded into Unreal Engine.

    \begin{figure}
        \centering
        \includegraphics[width = 1\textwidth]{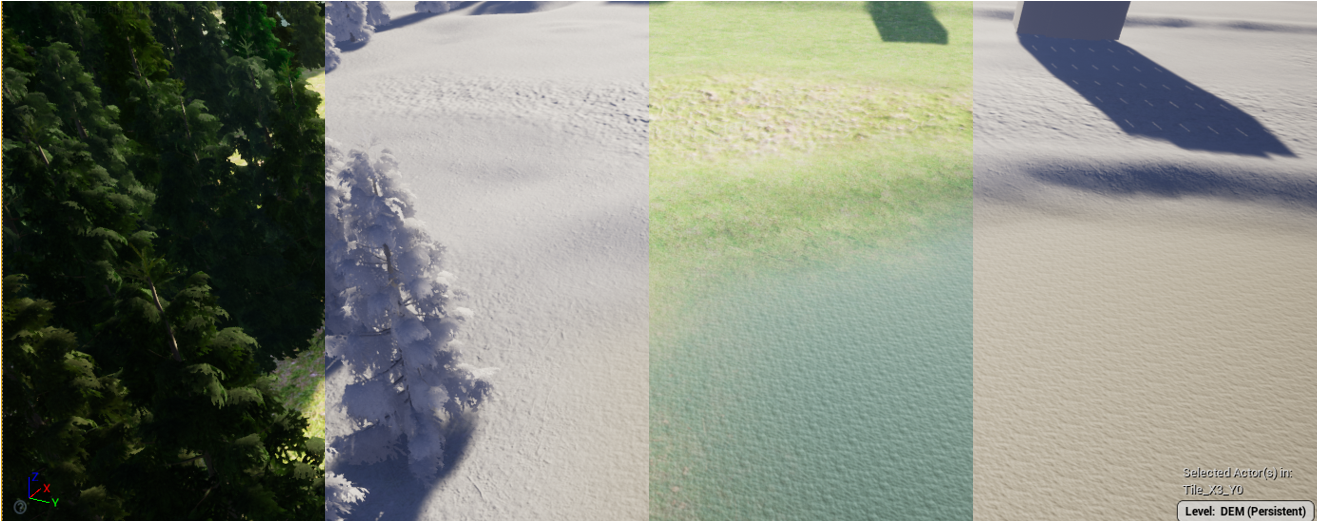}
        \caption{Composition of material-details in the 3D model - Unreal Engine.}
        \label{result_shaderdetails}
    \end{figure}
\begin{figure}
        \centering
        \includegraphics[width = 1\textwidth]{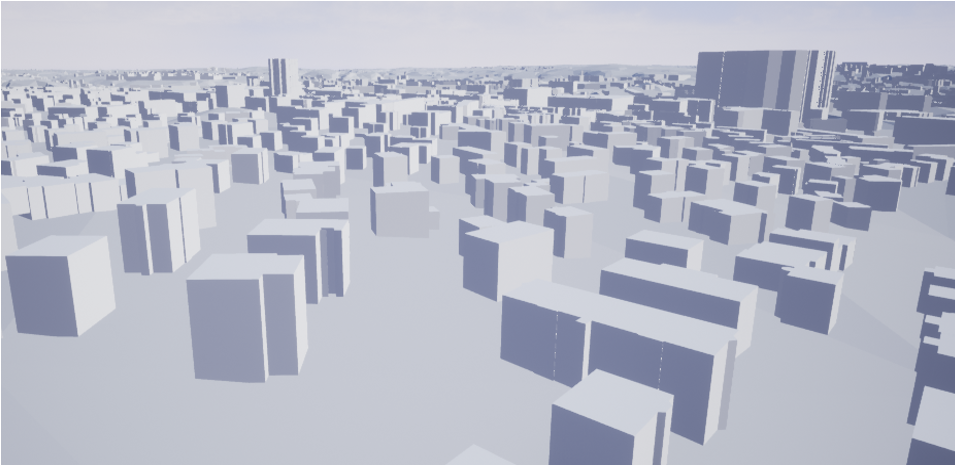}
        \caption{LoD1 Meshes within Unreal Engine - Datasmith, Unreal Engine.}
        \label{result_buildings}
    \end{figure}

In its material editor, Unreal Engine offers a "Landscape Layer Blend" node. The node enables the blending of multiple textures and materials linked to the landscape masks. A layer is added for each class of landscape material, and a texture is assigned to it.
Combining multiple layers allows adding a level of detail to the geometry without increasing the complexity of the 3D model. This is done by layering specific textures that inform Unreal Engine on how light interacts with these materials. Figure \ref{result_shaderdetails} shows an image from the final model with the diffuse colors turned on and off. In the sections where the colors are turned off, we see fine details on the ground that vary depending on the land-use type. These details are created using \textit{bump} and \textit{displacement} textures that contain information on how the smooth terrain can be distorted to provide natural imperfections to the model without requiring additional polygons. The buildings are loaded into Unreal Engine as LoD1 building meshes and then aligned to their coordinate position calculated in FME (see Figure \ref{result_buildings}).
    s

\subsection{Data Visualization}
\subsubsection{Isoline data}
\begin{figure}[H]
        \centering
        \includegraphics[width = 1\textwidth]{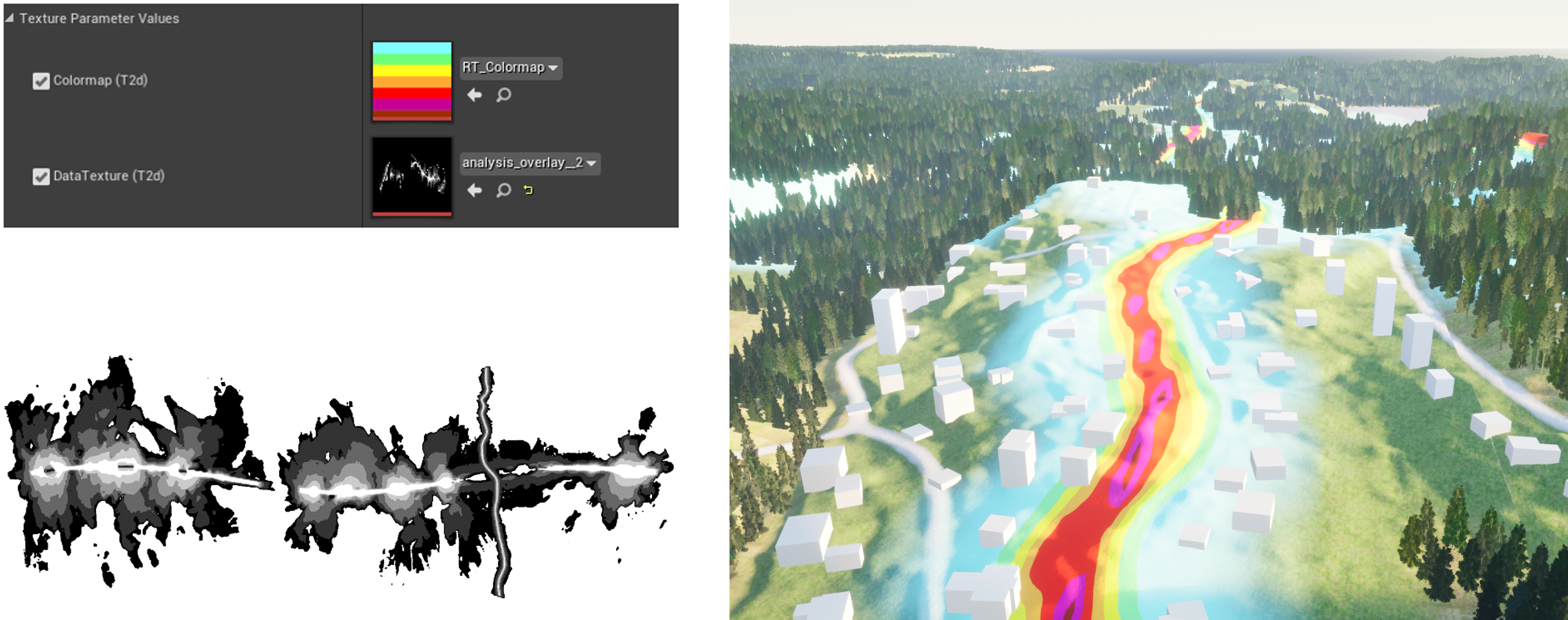}
        \caption{(Left-Top) Unreal Engine User interface for loading the color scale image and the grey scale mask of data to be visualized. (Left Bottom) Grey-scale mask to be visualized. (Right) Final visualization of grey scale mask - Unreal Engine.}
        \label{result_isoline}
    \end{figure}

A dataset containing the results of a noise simulation for a region was used to illustrate the visualization of isoline data. The data set contains vector data in the form of closed polygons representing different intensities of noise levels. The vector data is encoded into a 2D texture. The 2D texture is then remapped within Unreal Engine to a color scale selected by the user and draped over the landscape. Figure \ref{result_isoline} shows the results of visualizing the isoline data.
\subsubsection{Volumetric data}

\begin{figure}[H]
    \centering
    \includegraphics[width=\linewidth,height=\textheight,keepaspectratio]{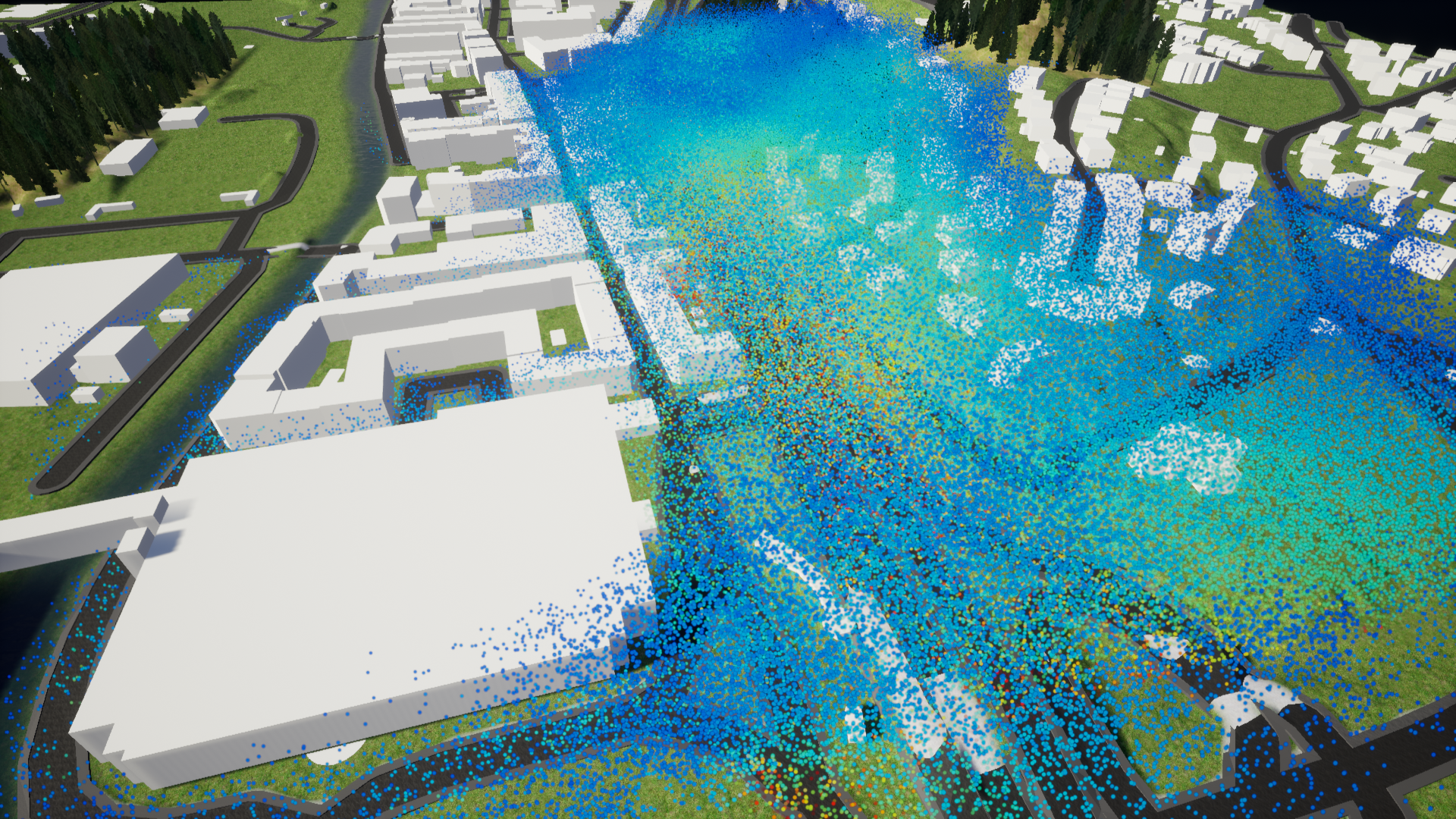}
    \caption{Volumetric data visualized with particles on top of generated building and terrain geometry. With up to 2 million particles, colored to show NO2 concentration and moving to visualize wind. The median frame time is 7 ms, with approximately 5 ms for the environment and 2 ms for the particle visualization.}
    \label{volumetric-data-vis}
\end{figure}

A dataset covering a volume of 512x512x64 was encoded into a data texture by folding the z-coordinate into an 8x8 grid in a 2D texture. This requires the total resolution of the texture to be (512x8)x(512x8) = 4196x4196. The calculations required to look up the data in this texture are done directly in a Niagara particle system module which can run on the GPU. Particles can be spawned and placed within this volume in several ways, but we primarily spawn particles randomly within the volume and kill them if they are, e.g., below a set threshold value. This approach makes it easy to vary the number of particles used to visualize the data depending on performance requirements.

\subsubsection{Streamlines}

\begin{figure}
    \centering
    \includegraphics[width=\linewidth,height=\textheight,keepaspectratio]{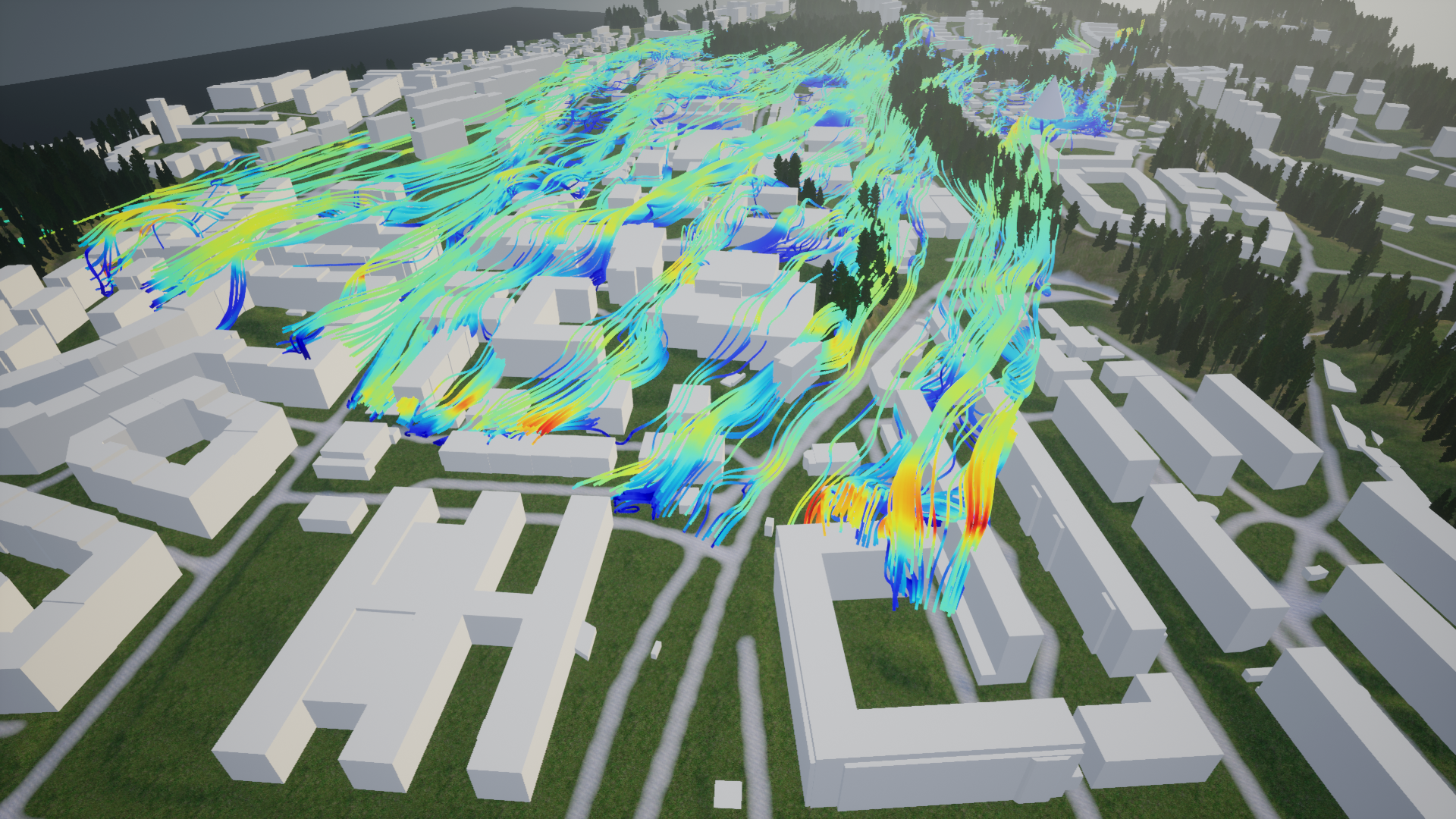}
    \caption{Streamlines overview showing 1003 (from a specific dataset) streamlines over generated building and landscape geometry.}
    \label{streamlines_overview}
\end{figure}

\begin{figure}[H]
    \centering
    \includegraphics[width=\linewidth,height=\textheight,keepaspectratio]{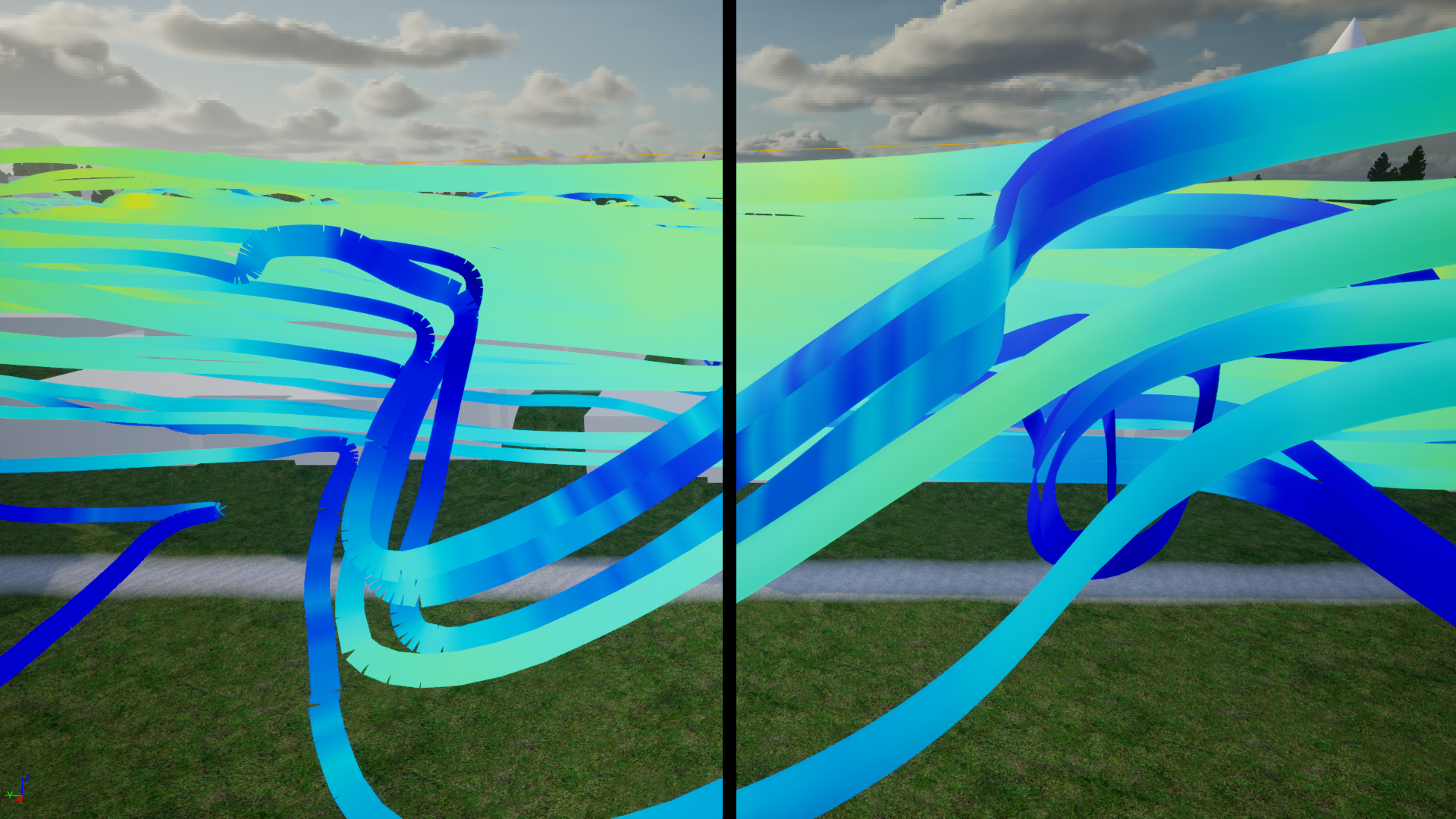}
    \caption{Streamlines comparison. (Left) displays the particle system implementation (Right) displays streamlines with procedural mesh.}
    \label{streamlines_comparison}
\end{figure}

Procedurally generated streamlines and particle streamlines have complementary values, as procedurally generated streamlines have higher visual quality up close, but particle streamlines perform better with many streamlines. Figure \ref{streamlines_overview} shows an overview at a distance where it is not possible to tell the difference visually. Here, it would generally be preferable to use particle streamlines. Figure \ref{streamlines_comparison} shows a close-up comparison, where artefacts can be seen in the sharp turns of particle streamlines to the left. Table 2 lists performance numbers for the two options with different numbers of streamlines.

\begin{center}
\begin{table}
\begin{tabular}{c c c} 
\toprule
 Streamlines Number & Procedural Streamlines & Particle Streamlines \\ [0.5ex] 
\cmidrule(l){1-3} 
 1,003 & 12 ms & 5 ms \\ 
 \hline
 4,158 & 47 ms & 12 ms \\
 \hline
 8,316 & 134 ms & 51 ms \\
 \hline
 12,474 & 247 ms & 127 ms \\

\end{tabular}
\caption{Comparison of performance of procedural streamlines and particle streamlines, with different numbers of streamlines. Median frame timings over 15 seconds on Nvidia GeForce RTX 2080 Ti.}
\end{table}
\end{center}


\section{Discussions and conclusion}
Generating virtual environments that represent the physical world is time-consuming and often complex. This paper presents a workflow for the procedural generation of such virtual worlds using commonly available GIS data sets enabled through modern game engines. In addition to generating virtual worlds, we present workflows to visualize data from large-scale analysis of the built environment in raster and vector formats. The methods presented in the paper can be elaborated to include a greater level of detail, such as the number of land use classes included in the model, increasing the point in the material and texture composition, and increasing the level of detail in the building meshes.
The results of the proposed workflow show how a realistic virtual city model can be generated and how large-scale data can be visualized efficiently and accurately for various data types.
One of the main advantages of our proposed workflow is that it allows decision-makers to view alternate scenarios, both real and simulated, to make well-informed decisions. For instance, decision-makers can use the resulting virtual city model to test different traffic scenarios and assess air pollution or noise levels resulting from a new construction project. Furthermore, our workflow enables data visualization in a way that provides a realistic context, allowing decision-makers to understand the implications of their decisions better.
However, there are limitations to the proposed method as well. Spatial data is often not available in all regions and, when available, may not be of high quality. Additionally, spatial data may be re-projected to a coordinate reference system. In contrast, the Unreal Engine environment and many programs used to generate the analysis results operate on a Cartesian coordinate system with different points of origin. This causes inconsistencies in the alignment of external meshes, such as the buildings or other analysis layers. Another limitation of our workflow is that it requires significant computational resources and GIS and game development expertise. Finally, the workflow we present is currently limited to the visualization of static data, and further research is required to enable the visualization of large-scale real-time urban data.

In conclusion, the workflow presented in this paper shows an innovative approach to creating 3D virtual city models and visualizing large-scale urban data using GIS, and leveraging the advances in real-time game engines. Viewing urban data in a realistic context can help decision-makers make more informed decisions. However, our workflow requires significant computational resources and expertise, and the availability and accuracy of GIS data needed for this workflow may be a limiting factor.

\section{Acknowledgements}
This work is part of the Digital Twin Cities Centre supported by Sweden’s Innovation Agency
Vinnova under Grant No. 2019-00041.

\section{Data sources}

All data used in the present study are public data obtained from the Swedish Mapping, Cadastral, and Land Registration Authority\footnote{\url{https://www.lantmateriet.se/}}. In particular, we have used the data set "Lantmäteriet: Laserdata NH 2019" in LAZ format for point clouds and property maps, the data collection "Lantmäteriet: Fastighetskartan bebyggelse 2020" in SHP format for property maps. Both data sets are in the EPSG:3006 coordinate system.

\subsection*{Implementation}
The algorithms described in this paper are implemented as part of the open-source (MIT license) Digital Twin Cities Platform~\cite{loggDigitalTwinCities2021} developed at the Digital Twin Cities Centre\footnote{\url{https://dtcc.chalmers.se/}}. The algorithms are implemented in Python, C++ based, and use several open-source packages, notably FEniCS~\cite{loggAutomatedSolutionDifferential2012} for sl differential equations expertise, Triangle~\cite{shewchukTriangleEngineering2D1996} for 2D mesh generation, and GEOS~\cite{geoscontributorsGEOSCoordinateTransformation2021} for any related geometrical operations.

\section{Declaration of Generative AI and AI-assisted technologies in the writing process}
During the preparation of this work, the author(s) used ChatGPT in order to rewrite parts of the text. After using this tool/service, the author(s) reviewed and edited the content as needed and take(s) full responsibility for the content of the publication.

\bibliographystyle{abbrv}
\bibliography{paper}

\end{document}